# On the Solution Uniqueness of Data-Driven Modeling of Flexible Loads

Shuai Lu, Jiayi Ding, Mingji Chen, Wei Gu, *Senior Member*, *IEEE*, Junpeng Zhu,
Yijun Xu, *Senior Member*, Zhaoyang Dong, *Fellow*, *IEEE*, Zezheng Sun

***Abstract*—This letter first explores the solution uniqueness of the data-driven modeling of price-responsive flexible loads (PFL). The PFL on the demand side is critical in modern power systems. An accurate PFL model is fundamental for system operations. However, whether the PFL model can be uniquely and correctly identified from operational data remains unclear. To address this, we analyze the structural and practical identifiability of the PFL model, deriving the dataset condition that guarantees the solution uniqueness. Besides, we point out the practical implications of the results. Numerical tests validate this work.**

## I. Introduction

THE demand-side flexible resources, including adjustable loads, distributed energy resources, and virtual power plants, are increasingly important in modern power systems. They can usually be modeled as price-responsive flexible loads (PFL) that actively respond to the electricity price to facilitate the system analysis. The model of the PFL that can accurately describe the temporal evolution of the aggregated power serves a critical interface role in power system operations.

Currently, two primary approaches exist for modeling the PFL: the physics-based and data-driven ones. The former starts from the individual components within the PFL. It usually frames the modeling of the PFL as a feasible region projection problem, and the commonly used techniques include the Minkowski sum methods [1], optimization-based outer/inner approximation methods [2], and heuristics-based feasible region elimination/expansion methods [3, 4]. However, this approach suffers from high computational demands and low accuracy, particularly as time periods expand. More critically, obtaining detailed models of each component within the PFL is impractical for distribution system operators (DSO).

Alternatively, the data-driven approach identifies the aggregated power model of PFL from the operational data, including the price and aggregated power, offering a more flexible and adaptable way for the DSO. This approach usually resorts to the inverse optimization (IO) technique, as the aggregated power of the PFL is determined by an optimization model parametrized by the price. To name a few, Tan *et al.* [5] prescribe a physics-informed parametric virtual battery to describe the aggregated power of the PFL, the parameters of which are then identified by the IO method. Lyu *et al.* [6] further improve this approach by using machine learning to enhance the parameter updating efficiency. While the above work establishes the basic framework, some fundamental problems remain unsolved. This letter focuses on one problem: whether the PFL model can be uniquely identified from the data. This problem is significant because if the solution uniqueness of the identification model cannot be guaranteed, the obtained PFL model may be inconsistent with reality, which will bring risks to power systems' operational safety and reliability.

To address this, we analyze the structural and practical identifiability of the PFL model and derive the condition of the dataset under which the PFL model can be uniquely identified. Then, the practical implications of the results are discussed. Finally, numerical simulation validates the effectiveness of the results.

## II. Problem Statement

Denote the decision period of the PFL as $T$. We assume that the model used by the PFL to determine the optimal aggregated power $P_*$ is as [5]:

$$P_* = \underset{P\in\Omega}{\operatorname{argmin}}\ \tilde{\lambda}^\top P, \tag{1}$$

wherein $P \in \mathbb{R}^T$ is the aggregated power variable; $\Omega$ is the feasible region of $P$ , i.e., the PFL model; and $\tilde{\lambda} \in \mathbb{R}^T$ is the PFL's predicted electricity prices before making decisions.

Based on the model (1), the DSO can estimate the optimal aggregated power $\hat{P}_* \in \mathbb{R}^T$ of the PFL under the price $\lambda$ by:

$$\mathrm{M}(\Omega):\quad \hat{P}_*(\lambda,\Omega) = \underset{P\in\Omega}{\operatorname{argmin}}\ \lambda^\top P, \tag{2}$$

where $\hat{P}_*$ is modeled as a function of $\lambda$ instead of $\tilde{\lambda}$, as the DSO cannot know the PFL's prediction $\tilde{\lambda}$.

In model (2), the feasible region $\Omega$ is usually characterized by the parameters $\theta$. A typical PFL model widely used in existing work is the storage-like model [4, 5], a physical model using the battery to simulate the behaviors of PFL, as:

$$\Omega(\theta) \triangleq \Omega_{phy}(\theta) = \left\{ P \left| \begin{array}{l} P = \sum_{n\in \mathrm{N}_{vb}} P_{vb}^n + \sum_{n\in \mathrm{N}_{td}} P_{td}^n + \sum_{n\in \mathrm{N}_{fix}} P_{fix}^n \\ \underline{E}_{vb}^n \le \Upsilon_1^n P_{vb}^n \Delta t + E_{vb,0}^n \Upsilon_2^n \le \overline{E}_{vb}^n, \\ \underline{P}_{vb}^n \le P_{vb}^n \le \overline{P}_{vb}^n, \forall n \in \mathrm{N}_{vb} \\ \underline{P}_{td}^n \le P_{td}^n \le \overline{P}_{td}^n, \forall n \in \mathrm{N}_{td} \end{array} \right. \right\}, \tag{3}$$

wherein $P_{vb}^n, P_{td}^n \in \mathbb{R}^T$ are the power of the temporally coupled (i.e., virtual storage-like)/temporally decoupled adjustable loads; $P_{fix}^n \in \mathbb{R}^T$ is the fixed load; $\underline{P}_{vb}^n, \overline{P}_{vb}^n \in \mathbb{R}^T$ are the lower and upper bounds of $P_{vb}^n$; $\underline{P}_{td}^n, \overline{P}_{td}^n \in \mathbb{R}^T$ are the lower and upper bounds of $P_{td}^n$; $\underline{E}_{vb}^n, \overline{E}_{vb}^n \in \mathbb{R}^T$ and $E_{vb,0}^n \in \mathbb{R}$ are the lower bound, upper bound, and initial value of the energy; $\Upsilon_1^n \in \mathbb{R}^{T\times T}$ is a lower triangular matrix, wherein $(\Upsilon_1^n)_{ij} = (\sigma^n)^{i-j}, i \ge j$ wherein $\sigma^n$ denotes the energy loss ratio; $\Upsilon_2^n \in \mathbb{R}^T$ and $(\Upsilon_2^n)_i = (\sigma^n)^i$; and $\mathrm{N}_{vb}/\mathrm{N}_{td}/\mathrm{N}_{fix}$ is the index set of the temporally coupled adjustable loads/temporally decoupled adjustable loads/fixed loads in the PFL. In this model, the model parameters are $\theta_{vb}^n \triangleq \left\{\underline{P}_{vb}^n, \overline{P}_{vb}^n, \underline{P}_{td}^n, \overline{P}_{td}^n, \underline{E}_{vb}^n, \overline{E}_{vb}^n, E_{vb,0}^n, \sigma^n\right\}$.

Since the PFL model is embedded in the optimization model (2), its identification is a typical IO problem. After collecting the price-power pairs $\left(\lambda^k, \tilde{P}_*^k\right), \forall k \in \mathrm{K}$, the DSO can use the following model to estimate the parameters $\theta$, as

$$\begin{aligned} \hat{\theta} = \arg\min_{\theta}\ & f(\theta) = \frac{1}{|\mathrm{K}|} \sum_{k\in\mathrm{K}} \left\| \tilde{P}_*^k - \hat{P}_*^k\left(\lambda^k, \Omega\right) \right\| \\ s.t.\ & \hat{P}_*^k\left(\lambda^k, \Omega\right) = \underset{P\in\Omega(\theta)}{\operatorname{argmin}}\ \left(\lambda^k\right)^\top P, \quad \forall k \in \mathrm{K} \end{aligned}, \tag{4}$$

wherein the superscript $k$ denotes the sample index, $\mathrm{K}$ denotes the sample set; $\hat{\theta}$ is the optimal estimated value of the model parameter $\theta$; $\hat{P}_*^k$ is the estimated aggregated power; and $\tilde{P}_*^k \in \mathbb{R}^T$ is the measurement of the aggregated power $P_*^k$, which can be modeled as $\tilde{P}_*^k = P_*^k + e_P^k$, wherein $e_P^k \in \mathbb{R}^T$ denotes the measurement errors of $P_*^k$.

The model (4) is the key problem of the data-driven modeling of the PFL. Note that the solution uniqueness of $\Omega$ in the model (4) is critical for the DSO because $\Omega$ describes the power adjustable ranges of the PFL, a false estimation of which will threaten the operational safety and reliability. Theoretically, if the model $\Omega(\theta)$ is well defined so that $\exists\theta$ making $\Omega(\theta) \neq \emptyset$, the model (4) always has (at least) a solution $\theta$ such that $f(\theta) < +\infty$. Under the conditions that $\lambda^k = \tilde{\lambda}^k, e_P^k = 0, \forall k \in \mathrm{K}$ and the model $\Omega(\theta)$ is correctly selected, the models (1) and (2) are consistent, i.e., the model (2) can accurately estimate the aggregated power $P_*$ in model (1) so that $\hat{P}_*(\lambda, \Omega) = P_*$, and thus the optimal value of model (4) will be $f(\theta) = 0$.

Besides, we clarify the focuses of this work as follows:

(1) While how to efficiently solve the IO problem (4) remains an open problem, this work focuses on whether its solution is unique under the dataset $\left(\lambda^k, \tilde{P}_*^k\right), \forall k \in \mathrm{K}$.

(2) Although the physical model $\Omega_{phy}$ plays an important role in identifying the PFL, this work does not focus on how to choose it, and the model (3) is just an example.

(3) Currently, in most regions of the world, electricity prices are pre-released and will not be updated frequently, such as fixed or time-of-use prices, meaning that the PFL can know exact prices before making decisions, i.e., $\lambda^k = \tilde{\lambda}^k, \forall k$. For the case where the PFL needs to predict the prices, such as real-time prices released after use, $\lambda^k = \tilde{\lambda}^k, \forall k$ may not hold. This work only focuses on the former.

## III. Solution Uniqueness

First, let us introduce the definitions and assumptions to be used. These definitions are less rigorous but enough for this work.

***Definition 1*** *(Structural identifiability).* For the system $y = y(u, \omega)$, assuming that input $u$ and output $y$ are noise-free, it is said to be structurally identifiable if for all $\omega$ and $\tilde{\omega}$ in the parameter space, there exists $u$ in the input space U making that $y(u, \omega) = y(u, \tilde{\omega})$ holds only when $\omega = \tilde{\omega}$.

***Definition 2*** *(Practical identifiability).* For the system $y = y(u, \omega)$, given the available input $u \in \mathrm{U}$ and output $y$ (i.e., real and noisy), it is said to be practically identifiable if for all $\omega$ and $\tilde{\omega}$ in the parameter space, $y(u, \omega) = y(u, \tilde{\omega}), \forall u \in \mathrm{U}$ implies $gap(\omega - \tilde{\omega}) \leq \epsilon$, wherein $gap(\cdot)$ is a measure quantifying the uncertainties in the estimates, and $\epsilon$ is a sufficiently small positive number.

***Assumption 1.*** The set $\Omega$ is a deterministic nonempty bounded polyhedron.

***Assumption 2.*** The dataset $\left(\lambda^k, \tilde{P}_*^k\right), k \in \mathrm{K}$ are noise-free, i.e., $e_P^k = 0, \left(\lambda^k, \tilde{P}_*^k\right) = \left(\lambda^k, P_*^k\right), \forall k \in \mathrm{K}$.

***Remark 1.*** Some further explanations are as follows:

(1) Both structural and practical identifiability focus on whether we can uniquely determine the parameters $\omega$. The former is a necessary condition for the latter.

(2) With *Assumption 2*, the practical identifiability is equivalent to under the actually available $u \in \mathrm{U}$, for all $\omega$ and $\tilde{\omega}$ in the parameter space, if $y(u, \omega) = y(u, \tilde{\omega}), \forall u \in \mathrm{U}$ implies $\omega = \tilde{\omega}$.

Essentially, the solution uniqueness of the data-driven modeling of the PFL is equivalent to the practical identifiability of the response model $\mathrm{M}(\Omega)$ defined in (2) under the given dataset $\left(\lambda^k, P_*^k\right), \forall k \in \mathrm{K}$, also equivalent to *the practical identifiability (or solution uniqueness) of the set $\Omega$ in model (4) under the given dataset* $\left(\lambda^k, P_*^k\right), \forall k \in \mathrm{K}$. Note that the solution uniqueness of the set $\Omega$ is not necessarily equivalent to that of the parameters $\theta$ in model (4). For example, for $\Omega(\theta) \triangleq \{P | AP \leq b\}$, $\theta = \{A, b\}$, we have $\Omega(\theta) = \Omega(k\theta), \forall k > 0$, indicating that any parameters $k\theta, k > 0$ will produce the same $\Omega$. This makes the analysis of the solution uniqueness of the set $\Omega$ in model (4) very complicated. To deal with this problem, with *Assumption 1*, we recast $\Omega(\theta)$ into a vertex-based convex hull, as

$$\Omega(\theta) \triangleq \Omega_{vert}(\theta) = \left\{ P \middle| \begin{array}{l} P = \theta\zeta,\ 0_V \leq \zeta \leq 1_V, \\ 1_V^\top \zeta = 1,\ \zeta \in \mathbb{R}^V \end{array} \right\}, \tag{5}$$

wherein $V$ is the number of vertices; $\theta = [\theta^1, \cdots, \theta^V] \in \mathbb{R}^{T \times V}$; $\theta^v \in \mathbb{R}^T$ is the coordinate of the $v$th vertex; $0_V = [0, \cdots, 0]^\top \in \mathbb{R}^V$; and $1_V = [1, \cdots, 1]^\top \in \mathbb{R}^V$.

It is evident that $\Omega_{vert}(\theta) = \Omega_{vert}(\tilde{\theta})$ yields $\theta = \tilde{\theta}$. Therefore, the solution uniqueness of the set $\Omega_{vert}$ in the model (4) is equivalent to that of the model parameters $\theta$. Namely, by the transformation (5), we can avoid the influence from the prior physical model of PFL and thus focus on purely data-driven modeling problems. Next, we first analyze the structural identifiability of the model $\Omega_{vert}$.

***Theorem 1*** *(Structural identifiability of $\Omega_{vert}$).* Under *Assumptions 1*, the set $\Omega_{vert}$ in model (4) is structurally identifiable.

***Proof.*** This theorem can be derived from the supporting hyperplane theorem [7]. Since $\Omega_{vert}(\theta)$ is a polyhedron, the supporting hyperplane theorem ensures that for the $v$th vertex $\theta^v$, there exists some vector $\lambda_*^v \in \mathbb{R}^T$ such that $(\lambda_*^v)^\top \theta^v \leq (\lambda_*^v)^\top P, \forall P \in \Omega_{vert}(\theta)$, wherein "=" holds if and only if $P = \theta^v$. Hence, we have $\hat{P}_*(\lambda_*^v, \Omega_{vert}) = \theta^v$. This implies that the vertices $\theta^v, v = 1, \cdots, V$ can be uniquely determined by choosing a proper input $\lambda_*^v$. Hence, once $\hat{P}_*\left(\lambda_*^v, \Omega_{vert}(\theta)\right) = \hat{P}_*\left(\lambda_*^v, \Omega_{vert}(\tilde{\theta})\right)$, we can conclude $\theta^v = \tilde{\theta}^v$. ■

***Remark 2.*** It can be concluded from *Theorem 1* that for a bounded (physics-informed or not) model $\Omega(\theta)$, its vertices are structurally identifiable, although the parameters $\theta$ may have different solutions. This implies that we do not need to consider whether the parameters $\theta$ can be uniquely identified when choosing a priori bounded (physical) model of PFL for the identification.

The next critical problem is the practical identifiability of $\Omega_{vert}$ under the dataset $\left(\lambda^k, P_*^k\right), \forall k \in \mathrm{K}$. Note that if $\exists\theta$ such that $\hat{P}_*\left(\lambda^k, \Omega_{vert}(\theta)\right) = P_*^k, \forall k \in \mathrm{K}$, the optimal value of model (4) is $f(\theta) = 0$, but not vice versa. The reason is that the model (4) could yield multiple different $\Omega_{vert}$ satisfying $\hat{P}_*\left(\lambda^k, \Omega_{vert}(\theta)\right) = P_*^k$. Therefore, we analyze the solutions of model (4) in the following.

Define the price matrix $\Lambda \triangleq [\lambda^1, \cdots, \lambda^K]^\top$, the aggregated power matrix $\Gamma_* \triangleq [P_*^1, \cdots, P_*^K]$, the cost matrix $\Xi \triangleq [(\lambda^1)^\top P_*^1, \cdots, (\lambda^K)^\top P_*^K]^\top$, and the set $\Pi \triangleq \{P | \Lambda P \geq \Xi\}$, wherein $K = |\mathrm{K}|$. Then, the following theorem gives some insights into the solutions of model (4).

***Theorem 2*** *(Practical identifiability of $\Omega_{vert}$).* For the model (4), under *Assumptions 1 & 2*, we have:

(a) $\mathrm{Conv}(\Gamma_*) \subseteq \Pi$;

(b) Any $\theta_1$ satisfying $\mathrm{Conv}(\Gamma_*) \subseteq \Omega_{vert}(\theta_1) \subseteq \Pi$ makes $f(\theta_1) = 0$, i.e., it is one of the optima of model (4);

(c) Any $\theta_2$ not satisfying $\mathrm{Conv}(\Gamma_*) \subseteq \Omega_{vert}(\theta_2) \subseteq \Pi$ makes $f(\theta_2) > 0$, i.e., it is not the optimum of model (4).

***Proof.*** This proof mainly utilizes the optimality property of the solution.

(a) Since the pair $\left(\lambda^k, P_*^k\right), \forall k \in \mathrm{K}$ is the solution of model (2), we have $(\lambda^k)^\top P_*^i \geq (\lambda^k)^\top P_*^k, \forall i, k \in \mathrm{K}$. This yields $\Lambda P_*^i \geq \Xi, \forall i \in \mathrm{K}$, i.e., $P_*^i \in \Pi, \forall i \in \mathrm{K}$, and hence we have $\mathrm{Conv}(\Gamma_*) \subseteq \Pi$.

(b) For $\theta_1$ satisfying $\mathrm{Conv}(\Gamma_*) \subseteq \Omega_{vert}(\theta_1) \subseteq \Pi$, since $P_*^k \in \mathrm{Conv}(\Gamma_*), \forall k \in \mathrm{K}$ and $\mathrm{Conv}(\Gamma_*) \subseteq \Omega_{vert}(\theta_1)$, we have $P_*^k \in \Omega_{vert}(\theta_1)$. Since $\Omega_{vert}(\theta_1) \subseteq \Pi$, we have $\Lambda P \geq \Xi, \forall P \in \Omega_{vert}(\theta_1)$. This indicates that for $\forall P \in \Omega_{vert}(\theta_1)$, we have $(\Lambda)_k P \geq (\Xi)_k, \forall k \in \mathrm{K}$, i.e., $(\lambda^k)^\top P \geq (\lambda^k)^\top P_*^k$, $\forall k \in \mathrm{K}$. Hence, with the input $\lambda^k$, one solution of model (2) is $\hat{P}_*(\lambda^k, \Omega_{vert}(\theta_1)) = P_*^k, \forall k \in \mathrm{K}$. Hence, we have $f(\theta_1) = 0$. Note that $\forall \theta, f(\theta) \geq 0$, and thus $\theta_1$ is one of the optima of model (4).

(c) The situation that $\theta_2$ does not satisfy $\mathrm{Conv}(\Gamma_*) \subseteq \Omega_{vert}(\theta_2) \subseteq \Pi$ includes two cases: $\mathrm{Conv}(\Gamma_*) \nsubseteq \Omega_{vert}(\theta_2)$ and $\Omega_{vert}(\theta_2) \nsubseteq \Pi$. The first case will obviously make $f(\theta_2) > 0$. We prove the second case in the following. For $\theta_2$ satisfying $\Omega_{vert}(\theta_2) \nsubseteq \Pi$, we can always find a point $P_o \in \Omega_{vert}(\theta_2)$ and $P_o \notin \Pi$. Since $P_o \notin \Pi$, there exists at least one $j \in \mathrm{K}$ such that $(\Lambda)_j P_o < (\Xi)_j$, i.e., $(\lambda^j)^\top P_o < (\lambda^j)^\top P_*^j$. Since $P_*^j \in \Pi$, we have $\hat{P}_*\left(\lambda^j, \Omega_{vert}(\theta_2)\right) = P_o \neq P_*^j$. This implies $\left\|P_*^j - P_o\right\| > 0$, and thus $f(\theta_2) > 0$. Besides, *Theorem 2*. (b) indicates that $\exists \theta$ such that $f(\theta) = 0$. Hence, $\theta_2$ is not the optimum of model (4). ■

***Remark 3.*** *Theorem 2* gives some insightful conclusions, which are explained as follows:

(1) *Existence of solutions*. For model (4), there must exist at least one optimum $\theta_*$ such that $f(\theta_*) = 0$, and $\theta_*$ is the optimum if and only if $\mathrm{Conv}(\Gamma_*) \subseteq \Omega_{vert}(\theta_*) \subseteq \Pi$.

(2) *Uniqueness of solutions*. Denote $\Delta\Omega \triangleq \complement_\Pi\left(\mathrm{Conv}(\Gamma_*)\right)$, wherein $\complement_X(Y)$ denotes the complement of Y in X. The case $\Delta\Omega = \emptyset$ provides a certificate for the practical identifiability of the PFL under $\Omega_{vert}(\theta)$, i.e., the solution uniqueness of the data-driven modeling of PFL.

(3) *Information completeness of dataset*. The case $\Delta\Omega \neq \emptyset$ indicates that the information in $(\lambda^k, P_*^k), \forall k \in \mathrm{K}$ is incomplete. In this case, it is unknown whether (part of) $\Delta\Omega$ should be included in $\Omega_{vert}(\theta)$. The (part of) $\Delta\Omega$ could be practically infeasible for the PFL or practically feasible but not being activated by the prices $\lambda^k, \forall k \in \mathrm{K}$. Note that $\mathrm{Conv}(\Gamma_*)$ (or $\Pi$) grows (or shrinks) as the effective information in $\Gamma_*$ increases.

(4) *Computation of* $\Delta\Omega$. Although $\Delta\Omega$ is hard to calculate, judging if $\Delta\Omega = \emptyset$ is equivalent to checking whether $\Pi \subseteq \mathrm{Conv}(\Gamma_*)$, i.e., checking if each point in the set $\Pi$ is a feasible point of $\mathrm{Conv}(\Gamma_*)$. This can be formulated into (mixed-integer) linear programming problems (see [8] for details).

Furthermore, we analyze the solution of model (4) without *Assumption 2* to provide more practical conclusions.

***Theorem 3.*** For the model (4), under *Assumption 1* and the $l^2$-norm objective function $f(\theta)$, supposing that $\forall k \in \mathrm{K}$, $e_P^k \sim \mathcal{N}(0_T, \Sigma_P)$, wherein $\Sigma_P \in \mathbb{R}^{T\times T}$ is the covariance matrix, we have:

(a) $\lim_{|\mathrm{K}|\to+\infty}\left(\hat{\theta}_{noise} - \hat{\theta}_{nf}\right) = 0$;

(b) $\left(f\left(\hat{\theta}_{noise}\right) - f\left(\hat{\theta}_{nf}\right)\right)_K \xrightarrow{Probability} \mathrm{tr}(\Sigma_P)$;

wherein $\hat{\theta}_{noise}$ and $\hat{\theta}_{nf}$ denote the solutions of model (4) under noisy $\tilde{P}_*^k$ and noise-free $P_*^k, \forall k \in \mathrm{K}$, respectively.

The proof is given in supplementary material [8].

***Remark 4.*** *Theorem 3* indicates that without *Assumption 2*, the model (4) can still identify an accurate PFL model using a large enough dataset. In this case, the value of $f\left(\hat{\theta}_{noise}\right) - \mathrm{tr}(\Sigma_P)$ ($\approx f\left(\hat{\theta}_{nf}\right)$) provides a basis for determining if the physical model $\Omega_{phy}(\theta)$ is correctly selected.

## IV. Practical Implications

The above theoretical results reveal that the PFL model (specifically, the feasible region $\Omega$) is not necessarily identifiable under the given dataset, in which model (4) will produce an incorrect PFL model that is inconsistent with reality. This will bring potential security risks and economic losses to the system operation but has not been fully noticed. Our results provide two implications for this problem, as follows.

(1) Checking the practical identifiability of $\Omega$

The results in Section III indicate that checking the practical identifiability of $\Omega$ is essential to avoid false identification results. As shown in *Remark 3. (3)*, the practical identifiability of $\Omega$ depends on the information completeness of the dataset. Based on this, *Remark 3. (4)* also provides a prior method to check the information completeness, which we can use to check the practical identifiability of $\Omega$ in real-world applications. Note that this method relies on the noise-free assumption (i.e., *Assumption 2*). If the noise is non-negligible, we can use statistical methods, such as Bayesian inference, to posteriorly analyze the practical identifiability of $\Omega$, which is worthy of further study.

(2) Enhancing the practical identifiability of $\Omega$

As indicated in *Remark 3. (3)*, the practical unidentifiability of $\Omega$ originates from the information completeness of the dataset. This inspires us to enhance practical identifiability using two different approaches. The first approach is to collect more effective data. The second approach is to incorporate a priori physical knowledge of the PFL to eliminate the indetermination. The priori physical model, for example, $\Omega_{phy}(\theta)$ in model (3), offers a concise but interpretable description of the PFL, and reduces the requirements for data quality and completeness. Essentially, this is equivalent to choosing a priori structure for the parameter space. *Remark 2* points out that it is unnecessary to concern about whether the parameters $\theta$ in $\Omega_{phy}(\theta)$ can be uniquely identified when a bounded set $\Omega_{phy}(\theta)$ is selected. Besides, in the real-world applications, when $e_P^k \sim \mathcal{N}(0_T, \Sigma_P), \forall k \in \mathrm{K}$, *Theorem 3* and *Remark 4* can be used to judge the correctness of the physical model $\Omega_{phy}(\theta)$.

## V. Numerical Test

To validate the above results, we perform simulations on a hypothetical PFL consisting of a fixed load, a time-decoupled adjustable load, and four batteries. First, we randomly generate the electricity price samples and use model (2) to get the aggregated power of the PFL. Second, we choose the physical model $\Omega_{phy}$ defined in (3) and use model (4) to identify the parameters, in which $\sigma^n$ is prescribed. The length of the period $T$ is set to 2 for visualization. To solve the model (4), we use the KKT conditions and big-M method to convert it into a single-level mixed-integer linear programming problem to resort to off-the-shelf solvers. The solution method and the detailed parameters and codes are provided in [8].

First, we investigate the impact of the sample size on sets $\mathrm{Conv}(\Gamma)$ and $\Pi$, as given in Fig. 1 (a). Consistent with theoretical results, the sets $\mathrm{Conv}(\Gamma)$ (or $\Pi$) expands (or shrinks) as the sample size increases. Besides, the $\Delta\Omega$ is still nonempty under 200 samples, indicating insufficient information. This means the operational characteristic of the PFL cannot be uniquely determined only by the current dataset, inspiring us to embed prior physical knowledge or design specific price vectors to detect if some undetermined region is feasible for the PFL. For example, we can choose any price $\lambda \in \{\lambda | \lambda^\top P_1 \geq 0, \lambda^\top P_2 \geq 0\}$ to probe the undetermined region D in Fig. 1 (a), the detailed principle of which is provided in [8].

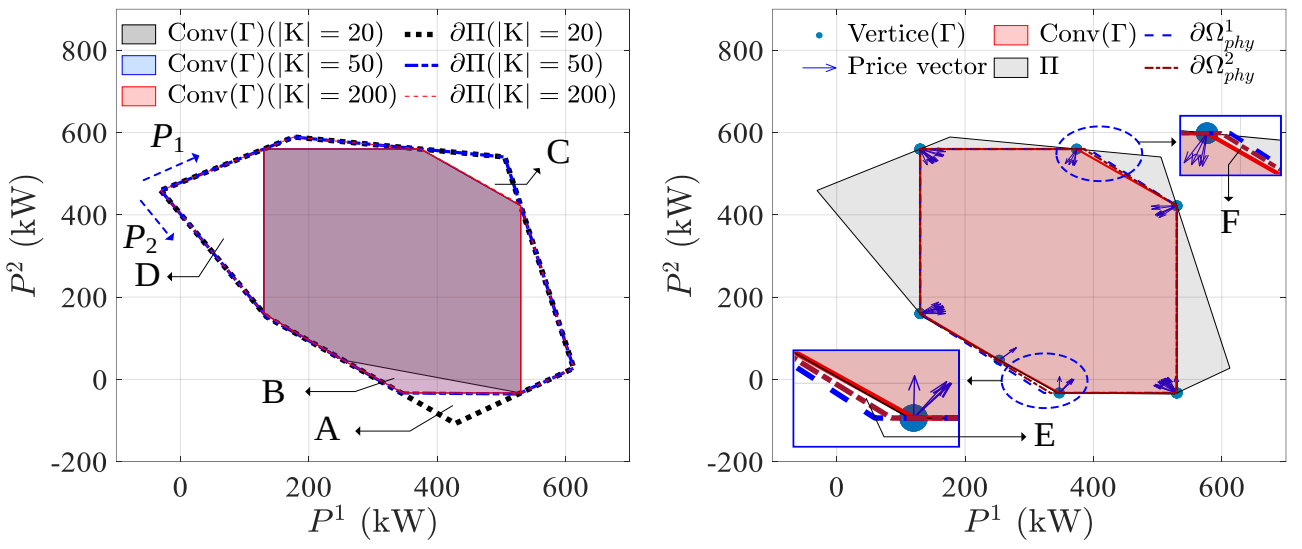

(a) (b)

Fig. 1. (a) Conv(Γ) and Π under different sample sizes; (b) Identification results of $\Omega_{phy}$ under different numbers of storage (1 and 2 for $\Omega_{phy}^1$ and $\Omega_{phy}^2$, respectively). (A: Π shrinks as the sample sizes |K| increases from 20 to 50; B: Conv(Γ) expands as |K| increases from 20 to 50; C: Π shrinks as the |K| increases from 50 to 200; D: Undetermined region; E: Practically infeasible region; F: Undetermined region in $\Omega_{phy}$.)

Second, we analyze the identification results using 50 samples based on the physical model $\Omega_{phy}$, in which the numbers of storage are set to 1 and 2 for $\Omega_{phy}^1$ and $\Omega_{phy}^2$, respectively. The feasible region of the aggregated power in $\Omega_{phy}$ is given in Fig. 1 (b). Obviously, the embedding of physical knowledge significantly eliminates the undetermined regions. Yet, the physical model $\Omega_{phy}$ in (3) is not exactly correct since it contains the practically infeasible region E. Interestingly, the undetermined region F is also identified, which is not covered by Conv(Γ). Besides, as the number of storage in the $\Omega_{phy}$ increases, the identification result is more accurate, consistent with the theoretical results.

In summary, the simulation results validate the theoretical results. Besides, we provide extended numerical test results in supplementary material [8].

## VI. Conclusion

This letter first investigates the solution uniqueness of the data-driven modeling of the PFL and gives some implications. We derive the condition of the dataset under which the PFL model can be uniquely identified from data. Overall, the data-driven modeling of the PFL is still in its initial stages. Future work includes (1) choosing a physics-compatible model for the PFL identification, (2) designing optimal prices to probe the undetermined region in the dataset, and (3) integrating the plug-and-play loads in the PFL.

# ■ Supplementary Material

*For the manuscript "Shuai Lu, Jiayi Ding, Mingji Chen, et al., On the Solution Uniqueness of Data-Driven Modeling of Flexible Loads".*

This material, as a supplement to the manuscript "*On the Solution Uniqueness of Data-Driven Modeling of Flexible Loads*", clarifies some fundamental problems and gives extended simulation results. Appendix A provides further explanations for *Remark 3. (4)*, showing how checking if $\Pi \subseteq \mathrm{Conv}(\Gamma_*)$ or not can be reformulated into (mixed-integer) linear programming problems. Appendix B provides the proof for *Theorem 3*. Appendix C introduces the solution method for the PFL identification model. Appendix D explains how to design price signals to probe the undetermined regions to handle the information insufficiency problem. Finally, Appendix E introduces the simulation settings and gives detailed simulation results.

## Appendix A
### CHECKING IF $\Pi \subseteq \mathbf{Conv}(\Gamma_*)$ OR NOT

In *Remark 3.* (4) in the manuscript, we discuss how to judge if $\Delta\Omega = \emptyset$, i.e., to check whether $\Pi \subseteq \mathrm{Conv}(\Gamma_*)$. As we analyzed in *Remark 3*, whether $\Delta\Omega = \emptyset$ holds depends on the information sufficiency of the dataset $(\lambda^k, P_*^k), k \in \mathrm{K}$, which is the endogenous property of the dataset. We said in the original manuscript that checking if $\Delta\Omega = \emptyset$ can be formulated into equivalent (mixed-integer) linear programming problems. Next, we will give more details about this problem.

First, let us recall the definition of $\Pi$ and $\Gamma_*$: $\Pi \triangleq \{P|\Lambda P \geq \Xi\}$, $\Gamma_* \triangleq [P_*^1 \cdots P_*^K]$, wherein $\Lambda \triangleq [\lambda^1 \cdots \lambda^K]^\top$ and $\Xi \triangleq [(\lambda^1)^\top P_*^1 \cdots (\lambda^K)^\top P_*^K]^\top$. According to this definition, after we collect the dataset (i.e., the price-power pairs) $(\lambda^k, P_*^k), k \in \mathrm{K}$, we can calculate the parameter matrix $\Lambda$ and parameter vector $\Xi$. To check whether $\Pi \subseteq \mathrm{Conv}(\Gamma_*)$ holds is equivalent to checking if $\forall P \in \Pi, P \in \mathrm{Conv}(\Gamma_*)$ holds.

Note that the set $\mathrm{Conv}(\Gamma_*)$ can be reformulated into an equivalent linear form by introducing auxiliary variables $\alpha = [\alpha^1 \cdots \alpha^K]^\top \in \mathbb{R}^K$ as:

$$\mathrm{Conv}(\Gamma_*) = \left\{P \middle| P = \Gamma_*\alpha, 1^\top\alpha = 1, 0 \leq \alpha \leq 1\right\}. \tag{A.1}$$

Then, to check if $\forall P \in \Pi, P \in \mathrm{Conv}(\Gamma_*)$ holds is equivalent to checking if $\forall P \in \Pi, \exists 0 \leq \alpha \leq 1, 1^\top\alpha = 1$ making $P = \Gamma_*\alpha$. For this problem, we can use two methods: vertex enumeration and robust optimization. The details are given as follows.

*(1) The vertex enumeration method*

In this method, we need to check if each vertex of the set $\Pi$ is in the set $\mathrm{Conv}(\Gamma_*)$. The corresponding mathematical model is

$$\begin{aligned}
&\forall P^j \in \mathrm{Vertex}(\Pi) = \{P^1, P^2, \cdots, P^J\}: \\
&\text{Solve feasibility-checking probelm}: \\
&\min_{P^j} \quad 0 \\
&s.t. \quad P^j = \Gamma_*\alpha \\
&\qquad\ \ 1^\top\alpha = 1 \\
&\qquad\ \ 0 \leq \alpha \leq 1
\end{aligned}, \tag{A.2}$$

wherein $\mathrm{Vertex}(\Pi)$ denotes the vertex set of the set $\Pi$.

If for each $P^j$, the above optimization model is always feasible, we can conclude that $\Pi \subseteq \mathrm{Conv}(\Gamma_*)$. For this method, we need to calculate the vertices of the set $\Pi$, and then check that each vertex is in the set $\mathrm{Conv}(\Gamma_*)$ by solving the feasibility checking problem (linear optimization problem) (A.2). If one of the feasibility-checking problems is infeasible, we can conclude that the corresponding vertex is not in the set $\mathrm{Conv}(\Gamma_*)$, i.e., $\Pi \nsubseteq \mathrm{Conv}(\Gamma_*)$.

Furthermore, we would like to clarify that this method is computationally expensive if the dimension of the set $\Pi$ is very high. Hence, it is unsuitable for the problem where the dimension of the set $\Pi$ is very high.

*(2) The robust optimization method*

Using the robust optimization method, the feasibility-checking problem can be formulated as a bilevel problem, as:

$$\max_{P\in\Pi} \min_{P_\# \in \mathrm{Conv}(\Gamma_*)} \quad \left\| P - P_\# \right\|_2^2 . \tag{A.3}$$

If the optimal objective of this model is 0, it means that $\forall P \in \Pi, P_\# \in \mathrm{Conv}(\Gamma_*)$ holds. This model can be further reformulated as:

$$\begin{aligned} &\max_{P\in\Pi} \min_{\alpha} \quad \left\| P - \Gamma_* \alpha \right\|_2^2 \\ &s.t. \quad 1^\top \alpha = 1 : \quad \mu \\ &\qquad\ \ 0 \le \alpha \le 1 : \quad \nu_1, \nu_2 \end{aligned} . \tag{A.4}$$

The above model can be equivalently transformed into a mixed-integer linear programming model using the KKT conditions and big-M method, as:

$$\begin{aligned} &\max_{P\in\Pi,\alpha,\mu,\nu_1,\nu_2,\varepsilon_1,\varepsilon_2} \quad 0 \\ &s.t. \quad 1^\top \alpha = 1 \\ &\qquad 0 \le \alpha \le 1 \\ &\qquad 2\Gamma_*^\top \left( \Gamma_* \alpha - P \right) + \mu \cdot 1 + \nu_1 - \nu_2 = 0 \\ &\qquad 0 \le \alpha \le \varepsilon_1 M \\ &\qquad -\left(1-\varepsilon_1\right) M \le \nu_1 \le 0 \\ &\qquad 0 \le 1 - \alpha \le \varepsilon_2 M \\ &\qquad -\left(1-\varepsilon_2\right) M \le \nu_2 \le 0 \\ &\qquad \varepsilon_1, \varepsilon_2 \in \{0,1\} \end{aligned} . \tag{A.5}$$

Based on the above, we show that checking if $\Delta\Omega = \emptyset$ can be formulated into equivalent (mixed-integer) linear programming problems.

## Appendix B

### PROOF FOR THEOREM 3

***Theorem 3.*** For the model (4), under *Assumption 1* and the $l^2$-norm objective function $f(\theta)$, supposing that $\forall k \in \mathrm{K}$, $e_P^k \sim \mathcal{N}(0_T, \Sigma_P)$, wherein $\Sigma_P \in \mathbb{R}^{T\times T}$ is the covariance matrix, we have:

(a) $\lim_{|\mathrm{K}|\to+\infty} \left( \hat{\theta}_{noise} - \hat{\theta}_{nf} \right) = 0$;

(b) $\left( f\left(\hat{\theta}_{noise}\right) - f\left(\hat{\theta}_{nf}\right) \right)_K \xrightarrow{Probability} \mathrm{tr}(\Sigma_P)$;

wherein $\hat{\theta}_{noise}$ and $\hat{\theta}_{nf}$ denote the solutions of model (4) under noisy $\tilde{P}_*^k$ and noise-free $P_*^k, \forall k \in \mathrm{K}$, respectively, and $\mathrm{tr}(\Sigma_P)$ denotes the trace of the matrix $\Sigma_P$.

***Proof.*** Recall that the model (4) in the manuscript is as follows:

$$\begin{aligned} &\hat{\theta} = \arg\min_{\theta} \ f(\theta) = \frac{1}{|\mathrm{K}|} \sum_{k\in\mathrm{K}} \left\| \tilde{P}_*^k - \hat{P}_*^k \left( \lambda^k, \Omega \right) \right\| \\ &s.t. \ \ \hat{P}_*^k \left( \lambda^k, \Omega \right) = \underset{P\in\Omega(\theta)}{\mathrm{argmin}} \ \left( \lambda^k \right)^\top P, \quad \forall k \in \mathrm{K} \end{aligned} , \tag{(4) in the manuscript}$$

wherein the superscript $k$ denotes the sample index, $\mathrm{K}$ denotes the sample set; $\hat{\theta}$ is the optimal estimated value of the model parameter $\theta$; $\hat{P}_*^k$ is the estimated aggregate power; and $\tilde{P}_*^k \in \mathbb{R}^T$ is the measurement of the aggregated power $P_*^k$, which can be modeled as $\tilde{P}_*^k = P_*^k + e_P^k$, wherein $e_P^k \in \mathbb{R}^T$ denotes the measurement errors of $P_*^k$.

We assume that $\forall k \in \mathrm{K},\ e_P^k \sim \mathcal{N}(\mu_P, \Sigma_P)$. Under *Assumption 1*, if the $l^2$-norm is used for $f(\theta)$, the model (4) can be rewritten as:

$$\begin{aligned} &\hat{\theta}_{noise} = \arg\min_{\theta} \quad f(\theta) = \frac{1}{|\mathrm{K}|}\sum_{k\in\mathrm{K}} \left\| \tilde{P}_*^k - \hat{P}_*^k\left(\lambda^k, \Omega\right)\right\|_2^2 \\ &s.t.\ \ \hat{P}_*^k\left(\lambda^k, \Omega\right) = \underset{P\in\Omega(\theta)}{\operatorname{argmin}} \ \left(\lambda^k\right)^\top P, \quad \forall k \in \mathrm{K} \end{aligned} . \tag{B.1}$$

Now, let us calculate the expression of $f(\theta)$, as:

$$\begin{aligned} f(\theta) &= \frac{1}{|\mathrm{K}|}\sum_{k\in\mathrm{K}} \left\| \tilde{P}_*^k - \hat{P}_*^k \right\|_2^2 \\ &= \frac{1}{|\mathrm{K}|}\sum_{k\in\mathrm{K}} \left\| P_*^k + e_P^k - \hat{P}_*^k \right\|_2^2 \\ &= \frac{1}{|\mathrm{K}|}\sum_{k\in\mathrm{K}} \left\| P_*^k - \hat{P}_*^k \right\|_2^2 + \frac{1}{|\mathrm{K}|}\sum_{k\in\mathrm{K}} \left\| e_P^k \right\|_2^2 + \frac{1}{|\mathrm{K}|}\sum_{k\in\mathrm{K}} 2\left(P_*^k - \hat{P}_*^k\right)^\top e_P^k \\ &= \frac{1}{|\mathrm{K}|}\sum_{k\in\mathrm{K}} \left\| P_*^k - \hat{P}_*^k \right\|_2^2 + \frac{1}{|\mathrm{K}|}\sum_{k\in\mathrm{K}} 2\left(P_*^k - \hat{P}_*^k\right)^\top \mu_P + \frac{1}{|\mathrm{K}|}\sum_{k\in\mathrm{K}} \left(\mu_P\right)^\top \mu_P \\ &\quad + \frac{1}{|\mathrm{K}|}\sum_{k\in\mathrm{K}} 2\left(P_*^k - \hat{P}_*^k\right)^\top \left(e_P^k - \mu_P\right) + \frac{1}{|\mathrm{K}|}\sum_{k\in\mathrm{K}} \left\| e_P^k \right\|_2^2 - \frac{1}{|\mathrm{K}|}\sum_{k\in\mathrm{K}} \left(\mu_P\right)^\top \mu_P \\ &= \frac{1}{|\mathrm{K}|}\sum_{k\in\mathrm{K}} \left\| P_*^k + \mu_P - \hat{P}_*^k \right\|_2^2 \\ &\quad + \frac{2}{|\mathrm{K}|}\sum_{k\in\mathrm{K}} \left(P_*^k - \hat{P}_*^k\right)^\top \left(e_P^k - \mu_P\right) + \frac{1}{|\mathrm{K}|}\sum_{k\in\mathrm{K}} \left\| e_P^k \right\|_2^2 - \left(\mu_P\right)^\top \mu_P \end{aligned} . \tag{B.2}$$

First, we analyze the term $\frac{2}{|\mathrm{K}|}\sum_{k\in\mathrm{K}}\left(P_*^k - \hat{P}_*^k\right)^\top\left(e_P^k - \mu_P\right)$. Note that the term $P_*^k - \hat{P}_*^k(\lambda^k, \Omega)$ is independent of the random error $e_P^k$, and $P_*^k - \hat{P}_*^k(\lambda^k, \Omega), \forall k \in \mathrm{K}$ can also be regarded as independent of each other. Since $\forall k \in \mathrm{K}, e_P^k \sim \mathcal{N}(\mu_P, \Sigma_P)$, we have $e_P^k - \mu_P \sim \mathcal{N}(0_T, \Sigma_P)$. Then, we have $\left(P_*^k - \hat{P}_*^k\right)^\top\left(e_P^k - \mu_P\right) \sim \mathcal{N}\left(0_T, \left(P_*^k - \hat{P}_*^k\right)^\top \Sigma_P \left(P_*^k - \hat{P}_*^k\right)\right)$. Thus, we have:

$$\begin{aligned} &\frac{2}{|\mathrm{K}|}\sum_{k\in\mathrm{K}} \left(P_*^k - \hat{P}_*^k\right)^\top \left(e_P^k - \mu_P\right) \sim \mathcal{N}\left(\mu_a, \sigma_a^2\right) \\ &\left(\mu_a = 0, \sigma_a^2 = \frac{4}{|\mathrm{K}|^2}\sum_{k\in\mathrm{K}} \left(P_*^k - \hat{P}_*^k\right)^\top \Sigma_P \left(P_*^k - \hat{P}_*^k\right)\right) \end{aligned} . \tag{B.3}$$

Without loss of generality, the vector $P_*^k - \hat{P}_*^k$ can be considered as bounded, i.e., there exits $M_a \in \mathbb{R}$ satisfying $\left|P_*^k - \hat{P}_*^k\right| \le M_a \cdot 1_T, \forall k \in \mathrm{K}$. Hence, we have:

$$\sigma_a^2 = \frac{4}{|\mathrm{K}|^2}\sum_{k\in\mathrm{K}} \left(P_*^k - \hat{P}_*^k\right)^\top \Sigma_P \left(P_*^k - \hat{P}_*^k\right) \le \frac{4}{|\mathrm{K}|^2}\sum_{k\in\mathrm{K}} M_a^2 \left(1_T\right)^\top \Sigma_P 1_T = \frac{4}{|\mathrm{K}|} M_a^2 \left(1_T\right)^\top \Sigma_P 1_T . \tag{B.4}$$

Taking the limit of $\sigma_a^2$, we have:

$$\lim_{|\mathrm{K}|\to+\infty} \sigma_a^2 \le \lim_{|\mathrm{K}|\to+\infty} \frac{4}{|\mathrm{K}|} M_a^2 \left(1_T\right)^\top \Sigma_P 1_T = 0 . \tag{B.5}$$

Hence, from Chebyshev's inequality, we have:

$$\lim_{|K|\to+\infty} P\left(\left|\frac{2}{|K|}\sum_{k\in K}\left(P_*^k-\hat{P}_*^k\right)^\top\left(e_P^k-\mu_P\right)\right|>\varepsilon\right)=0,\ \forall\varepsilon>0\,. \tag{B.6}$$

Second, we analyze the term $\frac{1}{|K|}\sum_{k\in K}\|e_P^k\|_2^2$. Since $e_P^k \sim \mathcal{N}(\mu_P,\Sigma_P)$, we have that $\forall k\in K$, $\|e_P^k\|_2^2$ follows the identical independent distribution. Then, based on the Central Limit Theorem, we have:

$$\begin{gathered}\frac{1}{|K|}\sum_{k\in K}\|e_P^k\|_2^2 \xrightarrow{\text{Distribution}} \mathcal{N}\left(\mu_b,\sigma_b^2\right)\\ \left(\mu_b=\operatorname{tr}\left(\Sigma_P\right)+\left(\mu_P\right)^\top\mu_P,\ \sigma_b^2=\frac{1}{|K|}\operatorname{var}\left(\|e_P^k\|_2^2\right)\right)\end{gathered}\,. \tag{B.7}$$

Note that

$$\begin{aligned}\operatorname{var}\left(\|e_p^k\|^2\right)&=\mathbb{E}\left(\|e_p^k\|^4\right)-\left[\mathbb{E}\left(\|e_p^k\|^2\right)\right]^2\\&\le\mathbb{E}\left(\|e_p^k\|^4\right)\\&\overset{\text{Cauchy inequality}}{\le}\mathbb{E}\left(T\sum_{t\in T}\left(\left(e_p^k\right)_t\right)^4\right)\\&\le T\sum_{t\in T}\left[3\left(\left(\Sigma_P\right)_{tt}\right)^4+6\left(\left(\Sigma_P\right)_{tt}\right)^2\left(\left(\mu_P\right)_t\right)^2+\left(\left(\mu_P\right)_t\right)^4\right]\end{aligned}\,. \tag{B.8}$$

We introduce a constant $M_b < +\infty$, defined as:

$$M_b=T\sum_{t\in T}\left[3\left(\left(\Sigma_P\right)_{tt}\right)^4+6\left(\left(\Sigma_P\right)_{tt}\right)^2\left(\left(\mu_P\right)_t\right)^2+\left(\left(\mu_P\right)_t\right)^4\right] \tag{B.9}$$

Then, we have:

$$\sigma_b^2\le\frac{1}{|K|}M_b \quad\text{and}\quad \lim_{|K|\to+\infty}\sigma_b^2\le\lim_{|K|\to+\infty}\frac{1}{|K|}M_b=0 \tag{B.10}$$

Hence, from Chebyshev's inequality, we have:

$$\lim_{|K|\to+\infty} P\left(\left|\frac{1}{|K|}\sum_{k\in K}\|e_P^k\|_2^2-\mu_b\right|>\varepsilon\right)=0,\ \forall\varepsilon>0\,. \tag{B.11}$$

Now, we define a random variable $\xi\in\mathbb{R}$, as:

$$\xi=\frac{2}{|K|}\sum_{k\in K}\left(P_*^k-\hat{P}_*^k\right)^\top\left(e_P^k-\mu_P\right)+\frac{1}{|K|}\sum_{k\in K}\|e_P^k\|_2^2-\left(\mu_P\right)^\top\mu_P-\operatorname{tr}\left(\Sigma_P\right). \tag{B.12}$$

Then, when $\mu_P = 0_T$, based on (B.2) and (B.12), the model (4) can be rewritten as:

$$\begin{gathered}\hat{\theta}_{noise}=\arg\min_{\theta}\ f(\theta)=\frac{1}{|K|}\sum_{k\in K}\left\|P_*^k-\hat{P}_*^k\left(\lambda^k,\Omega\right)\right\|_2^2+\operatorname{tr}\left(\Sigma_P\right)+\xi\\ s.t.\quad \hat{P}_*^k\left(\lambda^k,\Omega\right)=\operatorname*{argmin}_{P\in\Omega(\theta)}\ \left(\lambda^k\right)^\top P,\quad\forall k\in K\end{gathered}\,. \tag{B.13}$$

Compared with the case wherein *Assumption 2* holds, there are two extra terms in the objective function, including the constant $\operatorname{tr}(\Sigma_P)$ and the random variable $\xi$.

By substituting (B.6) and (B.11) into (B.12), we have:

$$\xi\xrightarrow{\text{Distribution}}\mathcal{N}\left(\mu_\xi,\sigma_\xi^2\right), \tag{B.14}$$

wherein the mean and variance can be calculated as:

$$\begin{cases} \mu_\xi = \mu_a + \mu_b - \text{tr}(\Sigma_P) - (\mu_P)^\top \mu_P = 0, \\ \sigma_\xi^2 = \sigma_a^2 + \sigma_b^2 = \frac{4}{|\mathrm{K}|^2} \sum_{k\in\mathrm{K}} \left(P_*^k - \hat{P}_*^k\right)^\top \Sigma_P \left(P_*^k - \hat{P}_*^k\right) + \frac{1}{|\mathrm{K}|} \text{var}\left(\left\|e_P^k\right\|_2^2\right) \end{cases}. \tag{B.15}$$

Based on this, we have:

$$\lim_{|\mathrm{K}|\to+\infty} \mu_\xi = 0, \lim_{|\mathrm{K}|\to+\infty} \sigma_\xi^2 = 0, \tag{B.16}$$

Hence, we have:

$$\lim_{|\mathrm{K}|\to+\infty} P\left(|\xi| > \varepsilon\right) = 0, \forall \varepsilon > 0, \tag{B.17}$$

Therefore, as $|\mathrm{K}| \to +\infty$, the random variable $\xi$ will not affect the solution of the model (B.13), i.e.,

$$\lim_{|\mathrm{K}|\to+\infty} \hat{\theta}_{noise} = \lim_{|\mathrm{K}|\to+\infty} \hat{\theta}_{nf}. \tag{B.18}$$

This proves *Theorem 3*. (a).

Further, note that:

$$f\left(\hat{\theta}_{nf}\right) = \frac{1}{|\mathrm{K}|} \sum_{k\in\mathrm{K}} \left\| P_*^k - \hat{P}_*^k\left(\lambda^k, \Omega\left(\hat{\theta}_{nf}\right)\right) \right\|_2^2, \tag{B.19}$$

and

$$\begin{aligned} f\left(\hat{\theta}_{noise}\right) - f\left(\hat{\theta}_{nf}\right) &= \frac{1}{|\mathrm{K}|} \sum_{k\in\mathrm{K}} \left\| P_*^k - \hat{P}_*^k\left(\lambda^k, \Omega\left(\hat{\theta}_{noise}\right)\right) \right\|_2^2 + \text{tr}(\Sigma_P) + \xi - f\left(\hat{\theta}_{nf}\right) \\ &= \frac{1}{|\mathrm{K}|} \sum_{k\in\mathrm{K}} \left\| P_*^k - \hat{P}_*^k\left(\lambda^k, \Omega\left(\hat{\theta}_{noise}\right)\right) \right\|_2^2 + \text{tr}(\Sigma_P) + \xi - \frac{1}{|\mathrm{K}|} \sum_{k\in\mathrm{K}} \left\| P_*^k - \hat{P}_*^k\left(\lambda^k, \Omega\left(\hat{\theta}_{nf}\right)\right) \right\|_2^2 \end{aligned}. \tag{B.20}$$

Based on (B.18), we have:

$$\begin{aligned} &\lim_{|\mathrm{K}|\to+\infty} P\left(\left| f\left(\hat{\theta}_{noise}\right) - f\left(\hat{\theta}_{nf}\right) - \text{tr}(\Sigma_P) \right| > \varepsilon\right) \\ &= \lim_{|\mathrm{K}|\to+\infty} P\left(\left| \frac{1}{|\mathrm{K}|} \sum_{k\in\mathrm{K}} \left\| P_*^k - \hat{P}_*^k\left(\lambda^k, \Omega\left(\hat{\theta}_{nf}\right)\right) \right\|_2^2 + \text{tr}(\Sigma_P) + \xi - f\left(\hat{\theta}_{nf}\right) - \text{tr}(\Sigma_P) \right| > \varepsilon\right) \\ &= \lim_{|\mathrm{K}|\to+\infty} \mathrm{P}\left(|\xi| > \varepsilon\right) \end{aligned}. \tag{B.21}$$

Combining (B.17) and (B.21), we have:

$$\lim_{|\mathrm{K}|\to+\infty} P\left(\left| f\left(\hat{\theta}_{noise}\right) - f\left(\hat{\theta}_{nf}\right) - \text{tr}(\Sigma_P) \right| > \varepsilon\right) = 0, \forall \varepsilon > 0. \tag{B.22}$$

This indicates that:

$$\left( f\left(\hat{\theta}_{noise}\right) - f\left(\hat{\theta}_{nf}\right) \right)_K \xrightarrow{\text{Probability}} \text{tr}(\Sigma_P). \tag{B.23}$$

This proves *Theorem 3*. (b). ■

From the above derivation, we can summarize the influence of noise on this identification model as follows:

(1) If the mean of measurement error $\mu_P$ is not 0, it will result in differing levels of drift in our estimated parameters, i.e., $\hat{\theta}_{noise} \neq \hat{\theta}_{nf}$. Fortunately, in practical engineering, the mean of the noise can be regarded as 0, i.e., $\mu_P = 0, \forall k$. Therefore, the estimated parameters $\hat{\theta}$ will not be affected when the sample size is large enough, i.e., $\hat{\theta}_{noise} = \hat{\theta}_{nf}$.

(2) While the constant term $\text{tr}(\Sigma_P)$ does not influence the optimal solution for $\theta$, it does affect the value of the objective function. Namely, the value of $f\left(\hat{\theta}_{noise}\right)$ will always be greater than 0, although the parameter $\theta$ is accurately estimated. Therefore, in this case, we do not have an exact certificate denoting that $\hat{\theta}_{noise}$ is the exact estimation, different from the noise-free case wherein $f\left(\hat{\theta}_{nf}\right) = 0$ indicates that $\hat{\theta}_{nf}$ is (one of) the optimal estimations. Still, a smaller $f(\hat{\theta}_{noise})$ means a better estimation of the parameters $\theta$. In this case, the value of $f(\hat{\theta}_{noise}) - \mathrm{tr}(\Sigma_P)$ $(\approx f(\hat{\theta}_{nf}))$ provides a basis for determining if the physical model $\Omega_{phy}(\theta)$ is correctly selected.

## Appendix C

### SOLUTION METHOD OF THE PFL IDENTIFICATION MODEL

Let us recall the PFL identification model (4) in the manuscript, as follows:

$$\begin{aligned} &\hat{\theta} = \arg\min_{\theta} \quad f(\theta) = \frac{1}{|\mathrm{K}|}\sum_{k\in\mathrm{K}} \left\| \tilde{P}_*^k - \hat{P}_*^k\left(\lambda^k, \Omega\right) \right\| \\ &s.t. \;\; \hat{P}_*^k\left(\lambda^k, \Omega\right) = \underset{P\in\Omega(\theta)}{\operatorname{argmin}} \;\; \left(\lambda^k\right)^\top P, \quad \forall k \in \mathrm{K} \end{aligned} , \qquad \text{((4) in the manuscript)}$$

wherein the superscript $k$ denotes the sample index, $\mathrm{K}$ denotes the sample set; $\hat{\theta}$ is the optimal estimated value of model parameter $\theta$; $\hat{P}_*^k$ is the estimated aggregate power; and $\tilde{P}_*^k \in \mathbb{R}^T$ is the measurement of the aggregated power $P_*^k$, which can be modeled as $\tilde{P}_*^k = P_*^k + e_P^k$, wherein $e_P^k \in \mathbb{R}^T$ denotes the measurement errors of $P_*^k$.

For the PFL identification model (4), if $\forall k \in \mathrm{K}$, $e_P^k = 0$ and the storage-like model $\Omega_{phy}$ is used, after the parameter $\sigma^n$ is prescribed, the above model can be reformulated as follows:

$$\begin{aligned} &\hat{\theta} = \underset{\theta=\left\{\underline{E}_{vb}^n, \bar{E}_{vb}^n, \underline{P}_{vb}^n, \bar{P}_{vb}^n, \underline{P}_{td}^n, \bar{P}_{td}^n\right\}}{\arg\min} \quad f(\theta) = \frac{1}{|\mathrm{K}|}\sum_{k\in\mathrm{K}} \left\| P_*^k - \hat{P}_*^k\left(\lambda^k, \Omega\right) \right\| \\ &s.t. \quad \forall k \in \mathrm{K}: \\ &\qquad \hat{P}_*^k\left(\lambda^k, \Omega\right) = \underset{P^k}{\operatorname{argmin}} \;\; \left(\lambda^k\right)^\top P^k \\ &\qquad s.t. \quad P^k = \sum_{n\in\mathrm{N}_{vb}} P_{vb}^{n,k} + \sum_{n\in\mathrm{N}_{td}} P_{td}^{n,k} + \sum_{n\in\mathrm{N}_{fix}} P_{fix}^{n,k} \quad : \quad \omega^k \\ &\qquad\qquad \underline{E}_{vb}^n \le \Upsilon_1^n P_{vb}^{n,k} \Delta t + E_{vb,0}^n \Upsilon_2^n \le \bar{E}_{vb}^n, \quad : \quad \nu_1^{n,k}, \nu_2^{n,k} \\ &\qquad\qquad \underline{P}_{vb}^n \le P_{vb}^{n,k} \le \bar{P}_{vb}^n, \forall n \in \mathrm{N}_{vb} \quad : \quad \nu_3^{n,k}, \nu_4^{n,k} \\ &\qquad\qquad \underline{P}_{td}^n \le P_{td}^{n,k} \le \bar{P}_{td}^n, \forall n \in \mathrm{N}_{td} \quad : \quad \nu_5^{n,k}, \nu_6^{n,k} \end{aligned} . \qquad \text{(C.1)}$$

The model (C.1) is a typical bilevel optimization problem, wherein the lower level is a linear optimization problem. Based on the duality theory, the lower-level model can be reformulated into KKT conditions, as:

$$\begin{aligned} \forall k \in \mathrm{K}: \;\; &\hat{P}_*^k = \sum_{n\in\mathrm{N}_{vb}} P_{vb}^{n,k} + \sum_{n\in\mathrm{N}_{td}} P_{td}^{n,k} + \sum_{n\in\mathrm{N}_{fix}} P_{fix}^{n,k} \\ &\underline{E}_{vb}^n \le \Upsilon_1^n P_{vb}^{n,k} \Delta t + E_{vb,0}^n \Upsilon_2^n \le \bar{E}_{vb}^n, \; \underline{P}_{vb}^n \le P_{vb}^{n,k} \le \bar{P}_{vb}^n, \forall n \in \mathrm{N}_{vb} \\ &\underline{P}_{td}^n \le P_{td}^{n,k} \le \bar{P}_{td}^n, \forall n \in \mathrm{N}_{td} \\ &\lambda^k + \omega^k = 0 \\ &-\omega^k + \left(\Upsilon_1^n\right)^\top \left(\nu_2^{n,k} - \nu_1^{n,k}\right)\Delta t - \nu_3^{n,k} + \nu_4^{n,k} = 0, \forall n \in \mathrm{N}_{vb} \\ &-\omega^k - \nu_5^{n,k} + \nu_6^{n,k} = 0, \forall n \in \mathrm{N}_{td} \\ &\nu_1^{n,k}, \nu_2^{n,k}, \nu_3^{n,k}, \nu_4^{n,k} \ge 0, \forall n \in \mathrm{N}_{vb} \\ &\nu_5^{n,k}, \nu_6^{n,k} \ge 0, \forall n \in \mathrm{N}_{td} \\ &\nu_1^{n,k} \odot \left(\underline{E}_{vb}^n - \Upsilon_1^n P_{vb}^{n,k} \Delta t - E_{vb,0}^n \Upsilon_2^n\right) = 0, \forall n \in \mathrm{N}_{vb} \\ &\nu_2^{n,k} \odot \left(\Upsilon_1^n P_{vb}^{n,k} \Delta t + E_{vb,0}^n \Upsilon_2^n - \bar{E}_{vb}^n\right) = 0, \forall n \in \mathrm{N}_{vb} \\ &\nu_3^{n,k} \odot \left(\underline{P}_{vb}^n - P_{vb}^{n,k}\right) = 0, \forall n \in \mathrm{N}_{vb} \\ &\nu_4^{n,k} \odot \left(P_{vb}^{n,k} - \bar{P}_{vb}^n\right) = 0, \forall n \in \mathrm{N}_{vb} \\ &\nu_5^{n,k} \odot \left(\underline{P}_{td}^n - P_{td}^{n,k}\right) = 0, \forall n \in \mathrm{N}_{td} \\ &\nu_6^{n,k} \odot \left(P_{td}^{n,k} - \bar{P}_{td}^n\right) = 0, \forall n \in \mathrm{N}_{td} \end{aligned} . \qquad \text{(C.2)}$$

Based on this, the model (C.1) using can be converted into a single-level optimization model, as:

$$\hat{\theta} = \underset{\substack{\theta=\{\underline{E}_{vb}^{n},\bar{E}_{vb}^{n},\underline{P}_{vb}^{n},\bar{P}_{vb}^{n},\underline{P}_{td}^{n},\bar{P}_{td}^{n}\},\\ \hat{P}_{*}^{k},P_{vb}^{n,k},P_{td}^{n,k},P_{fix}^{n,k},\\ \omega^{k},\nu_1^{n,k},\nu_2^{n,k},\nu_3^{n,k},\nu_4^{n,k},\nu_5^{n,k},\nu_6^{n,k}}}{\arg\min} \; f(\theta) = \underset{\substack{\theta=\{\underline{E}_{vb}^{n},\bar{E}_{vb}^{n},\underline{P}_{vb}^{n},\bar{P}_{vb}^{n},\underline{P}_{td}^{n},\bar{P}_{td}^{n}\},\\ \hat{P}_{*}^{k},P_{vb}^{n,k},P_{td}^{n,k},P_{fix}^{n,k},\\ \omega^{k},\nu_1^{n,k},\nu_2^{n,k},\nu_3^{n,k},\nu_4^{n,k},\nu_5^{n,k},\nu_6^{n,k}}}{\arg\min} \frac{1}{|\mathrm{K}|}\sum_{k\in\mathrm{K}} \left\| P_{*}^{k} - \hat{P}_{*}^{k} \right\| \quad s.t. \quad (\mathrm{C.2}) \quad . \tag{C.3}$$

By using the big-M method and introducing binary variables, the complementary constraints in (C.2) can be converted into mixed-integer linear programming, and thus, the model (C.3) can be finally converted into a mixed-integer linear optimization model for the $l^1$-norm objective function (or mixed-integer quadratic optimization model for the $l^2$-norm objective function).

Here, we clarify that the above method is not computationally efficient since many binary variables need to be introduced, especially when the sample size is large. Some more efficient solution methods have been investigated in [1, 2]. Since this work mainly focuses on whether the PFL model can be uniquely identified from the data, i.e., the solution uniqueness of the model (C.1), we do not go further into the solution method.

## Appendix D

### PRICE SIGNAL DESIGN FOR PROBING UNDETERMINED REGION

Denote $P_{\#}^{q}$ as the $q$th vertex of the set $\Pi$, $q = 1,2,\dots,Q_{\Pi}$. We assume that $P_{\#}^{j}$ is the vertex that is not in the set $\mathrm{Conv}(\Gamma)$. We need to collect more data to determine if $P_{\#}^{j}$ is in the feasible region $\Omega$ of the PFL. Specifically, we need to design a price signal $\lambda_{\#}$ to probe the response of the PFL to check if $P_{\#}^{j}$ is a feasible point for the PFL. In the following, we discuss two cases.

*(1) The first case: $P_{\#}^{j}$ is a feasible point of the set $\Omega$*

In this case, for the price signal $\lambda_{\#}$, we should have:

$$\lambda_{\#}: \quad P_{\#}^{j} = \underset{P\in\Omega}{\operatorname{argmin}} \; (\lambda_{\#})^{\top} P \; . \tag{D.1}$$

Since we do not know the exact $\Omega$ now, it is not easy to design the price signal $\lambda_{\#}$. However, since $\Omega$ is a subset of $\Pi$, we can use $\Pi$ to replace $\Omega$ in the above model to get a conservative signal, i.e.,

$$\lambda_{\#}: \quad P_{\#}^{j} = \underset{P\in\Pi}{\operatorname{argmin}} \; (\lambda_{\#})^{\top} P \; . \tag{D.2}$$

Note that the $\lambda_{\#}$ that makes the model (D.1) hold will naturally make the model (D.2) hold in this case. Therefore, we only need to analyze how to design $\lambda_{\#}$ for the model (D.2). The model (D.2) indicates:

$$(\lambda_{\#})^{\top} P_{\#}^{j} < (\lambda_{\#})^{\top} P_{\#}^{q} \quad \forall q = 1,\cdots,j-1,j+1,\cdots,Q_{\Pi} \quad . \tag{D.3}$$

This is equivalent to

$$(\lambda_{\#})^{\top} \left(P_{\#}^{j} - P_{\#}^{q}\right) < 0 \quad \forall q = 1,\cdots,j-1,j+1,,\cdots,Q_{\Pi} \quad . \tag{D.4}$$

Denoting $\Delta P_{\#}^{j,q} = P_{\#}^{q} - P_{\#}^{j}$, we have:

$$(\lambda_{\#})^{\top} \left(-\Delta P_{\#}^{j,q}\right) < 0 \quad \forall q = 1,\cdots,j-1,j+1,,\cdots,Q_{\Pi} \quad . \tag{D.5}$$

Note that since $\Pi$ is a polyhedron, and thus there exists $0 \le \alpha^{q,s} \le 1, \sum_{s\in S_j} \alpha^{q,s} = 1$ satisfying

$$\Delta P_{\#}^{j,q} = \sum_{s\in S_j} \alpha^{q,s} \Delta P_{\#}^{j,s}, \quad \forall q = 1,2,\cdots,Q_{\Pi} \,, \tag{D.6}$$

wherein $S_j$ is the set of the vertices connecting to the vertex $j$.

Substituting (D.6) into (D.5), we have:

$$\left(\lambda_{\#}\right)^{\top}\sum_{s\in S_j}\alpha^{q,s}\Delta P_{\#}^{j,s}>0\quad\forall q=1,\cdots,j-1,j+1,,\cdots,Q_{\Pi}\quad, \tag{D.7}$$

which is equivalent to the following:

$$\left(\lambda_{\#}\right)^{\top}\sum_{s\in S_j}\alpha^{s}\Delta P_{\#}^{j,s}>0\quad\forall\alpha^{s}:0\le\alpha^{s}\le 1,\sum_{s\in S_j}\alpha^{s}=1. \tag{D.8}$$

The above inequalities can be reformulated into

$$\left(\lambda_{\#}\right)^{\top}\Delta P_{\#}^{j,s}>0\quad\forall s\in S_j\cdot \tag{D.9}$$

Using the price signal satisfying the condition (D.9), if the vertex $P_{\#}^{j}$ is a feasible point of the $\Omega$, we will get a pair $\left(\lambda_{\#},P_{\#}^{j}\right)$ that will be added into the dataset. Then, $\mathrm{Conv}(\Gamma)$ will also contain $P_{\#}^{j}$.

Note that in our numerical test, since the dimension of the set $\Pi$ is 2, the formula (9) turns into:

$$\left(\lambda_{\#}\right)^{\top}P_1>0,\left(\lambda_{\#}\right)^{\top}P_2>0\cdot \tag{D.10}$$

*(2) The second case: $P_{\#}^{j}$ is not a feasible point of the set*

In this case, using the price signal satisfying the condition (D.9), the response of the PFL is assumed to be $\hat{P}_{\#}^{j}$, which is not $P_{\#}^{j}$. This indicates

$$\left(\lambda_{\#}\right)^{\top}\hat{P}_{\#}^{j}>\left(\lambda_{\#}\right)^{\top}P_{\#}^{j}\cdot \tag{D.11}$$

With this new pair $\left(\lambda_{\#},\hat{P}_{\#}^{j}\right)$, the set $\Pi$ is updated as:

$$\Pi_{\#}=\left\{P\middle|\left[\lambda^{1}\cdots\lambda^{K}\ \ \lambda_{\#}\right]^{\top}P\ge\left[\left(\lambda^{1}\right)^{\top}P_{*}^{1}\cdots\left(\lambda^{K}\right)^{\top}P_{*}^{K}\ \ \left(\lambda_{\#}\right)^{\top}\hat{P}_{\#}^{j}\right]\right\}\cdot \tag{D.12}$$

The formula (D.11) indicates that $[\lambda^{1}\ \cdots\ \lambda^{K}\ \ \lambda_{\#}]^{\top}P_{\#}^{j}\ge\left[(\lambda^{1})^{\top}P_{*}^{1}\ \cdots\ (\lambda^{K})^{\top}P_{*}^{K}\ \ (\lambda_{\#})^{\top}\hat{P}_{\#}^{j}\right]$ does not hold. Hence, $P_{\#}^{j}$ is not in the set of $\Pi_{\#}$. This means that if $P_{\#}^{j}$ is not a feasible point of the set $\Omega$, by adding the data pair consisting of the price signal $\lambda_{\#}$ and the corresponding response $\hat{P}_{\#}^{j}$, the new set $\Pi_{\#}$ will not contain $P_{\#}^{j}$.

In summary, the price signal $\lambda_{\#}$ can be used to probe the response of the PFL to check if $P_{\#}^{j}$ is a feasible point for the PFL and thus to reduce the set $\Delta\Omega$.

## Appendix E

## NUMERICAL TEST

*(1) Settings of tests*

To validate the results, we perform simulations on a hypothetical PFL consisting of a fixed load, a time-decoupled adjustable load, and four batteries. The length of the period is set to 2 for visualization. The detailed parameters and codes are provided at [3].

The simulation is divided into two steps, including:

*Step 1.* We generate a random set of electricity price samples for $\lambda\in\mathbb{R}^{N\times T}$. Then, the corresponding aggregate power $P_{*}\in\mathbb{R}^{N\times T}$ for is calculated by the response model (1). This process generates the dataset $(\lambda^{k},P_{*}^{k}),\forall k\in$ K , which is required for the identification of the PFL.

*Step 2.* We select the physical model $\Omega_{phy}$ parameterized by $\theta_{vb}^{n}$ as defined in (2), in which the number of the components, i.e., $\mathrm{N}_{vb}/\mathrm{N}_{td}/\mathrm{N}_{fix}$, are prescribed. Then, we substitute the dataset $(\lambda^{k},P_{*}^{k}),\forall k\in\mathrm{K}$ and the selected physical model $\Omega_{phy}$ into the identification model (4). Finally, we solve the identification model (4) to obtain the model parameters $\theta_{vb}^{n}=\left(\underline{P}_{vb}^{n},\overline{P}_{vb}^{n},\underline{P}_{td}^{n},\overline{P}_{td}^{n},\underline{E}_{vb}^{n},\overline{E}_{vb}^{n},E_{vb,0}^{n},\sigma^{n}\right)$. In the simulations, to facilitate the model solving, the value of the parameter $\sigma^{n}$ is prescribed.

After getting the model parameters $\theta_{vb}^{n}$, we can get the physical model $\Omega_{phy}$ by substituting the parameters. In the following analysis, the aggregated power feasible region of $\Omega_{phy}$ is compared with $\text{Conv}(\Gamma)$ and $\Pi$ to evaluate the accuracy of the identified results.

*(2) Test with noise-free data*

The numerical simulations in this section are based on noise-free data, i.e., we assume that the dataset $(\lambda^k, P_*^k), \forall k \in \mathrm{K}$ are noise-free. The identifications under different dataset sizes are tested.

First, we investigate the impact of the sample size on the sets $Conv(\Gamma)$ and $\Pi$, as given in Fig. E-1 (a). Consistent with theoretical results, the sets $\text{Conv}(\Gamma)$ (or $\Pi$) expands (or shrinks) as the sample size increases. For example, observing the four regions labeled A, B, C, and D in Fig. E-1 (a) and combining with Theorem 2, we can conclude that in the results under 20 samples, A, B, C, and D are all the undetermined regions since they are out of the set $Conv(\Gamma)$ but bounded by the boundary $\partial\Pi$. When the sample size increases from 20 to 50, the boundary $\partial\Pi$ shrinks so that the region A is not within the boundary $\partial\Pi$, indicating that the region A can be identified as the infeasible region of the PFL. At the same time, as the set $\text{Conv}(\Gamma)$ expands, the region B is contained within the set $Conv(\Gamma)$, and thus, it can be identified as the feasible region of the PFL. In the result under 200 samples, the boundary $\partial\Pi$ shrinks further and accordingly the region C is identified as an infeasible region. Besides, the $\Delta\Omega$ is still nonempty under 200 samples, indicating that the information in the dataset is insufficient. The theoretical results also show that as the sample size increases, the set $\Delta\Omega$ will gradually converge to 0, and finally, the exact feasible region can be uniquely determined. Obviously, the simulation results are in agreement with the theoretical ones. Furthermore, the results indicate that when $\Delta\Omega \neq \emptyset$, there always exist some undetermined regions that we cannot ensure whether they are part of the feasible region of the PFL. This inspires us to design specific price vectors to detect if some undetermined region is feasible for the PFL. For example, we can choose any price $\lambda \in \{\lambda | \lambda^{\top} P_1 \geq 0, \lambda^{\top} P_2 \geq 0\}$ to probe the undetermined region D in Fig. E-1 (a).

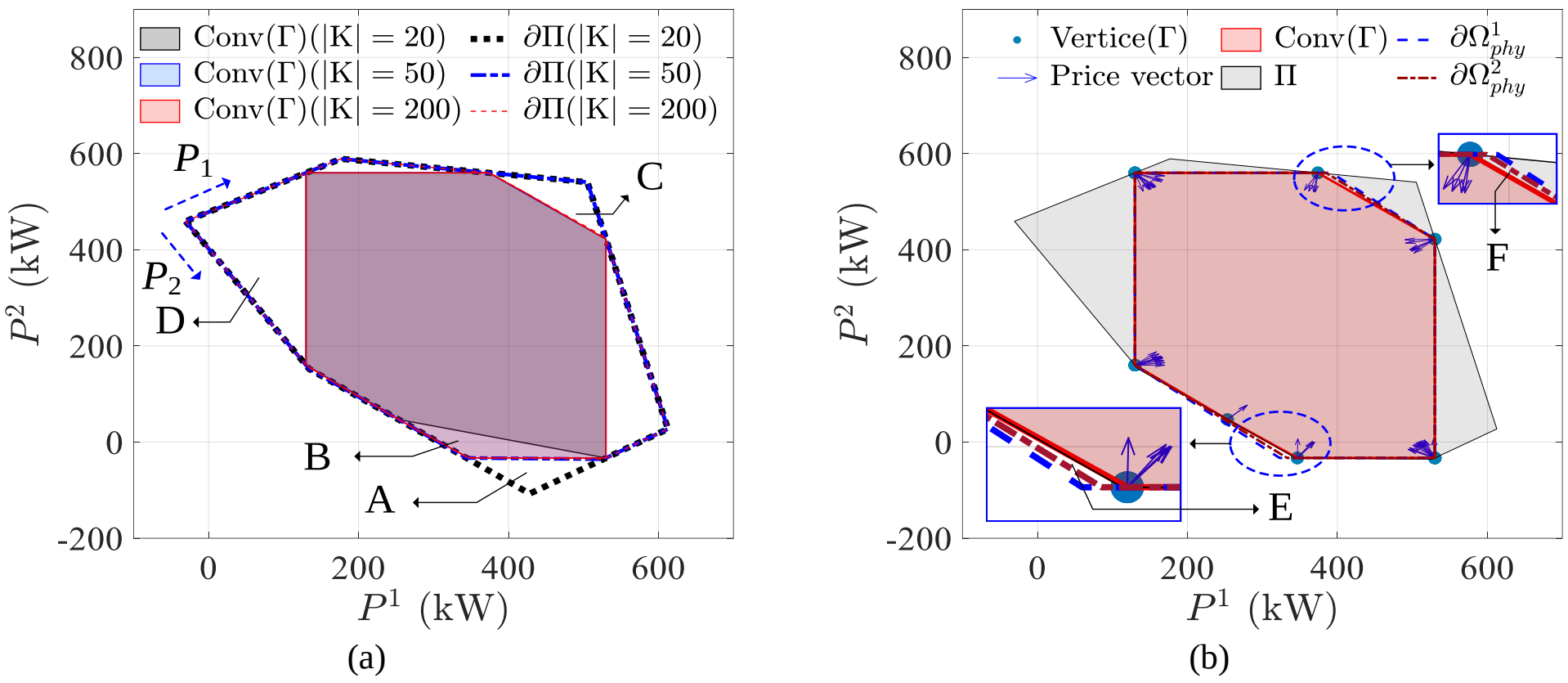

Fig. E-1. (a) $\text{Conv}(\Gamma)$ and $\Pi$ under different sample sizes; (b) Identification results of $\Omega_{phy}$ under different number of storage (1 and 2 for $\Omega_{phy}^1$ and $\Omega_{phy}^2$, respectively). (A: $\Pi$ shrinks as the sample sizes $|\mathrm{K}|$ increases from 20 to 50; B: $\text{Conv}(\Gamma)$ expands as $|\mathrm{K}|$ increases from 20 to 50; C: $\Pi$ shrinks as the $|\mathrm{K}|$ increases from 50 to 200; D: Undetermined region; E: Practically infeasible region; F: Undetermined region in $\Omega_{phy}$.)

Second, we analyze the identification results using 50 samples based on the physical model $\Omega_{phy}$, in which the numbers of storage are set to 1 and 2 for $\Omega_{phy}^1$ and $\Omega_{phy}^2$, respectively. The feasible region of the aggregated power in $\Omega_{phy}$ is given in Fig. E-1 (b). Obviously, the physical model $\Omega_{phy}$ in (2) is incorrect since it includes the practically infeasible region E that is outside of the set $\Pi$. Interestingly, the undetermined region F is also identified, which is not covered by $\text{Conv}(\Gamma)$. This indicates that neither the 50 samples nor the physical model $\Omega_{phy}$ includes the information that can determine whether the region F is part of the feasible region of PFL. Besides, it can be seen that $\Omega_{phy}^2$ is closer to $\text{Conv}(\Gamma)$ than $\Omega_{phy}^1$ since both the infeasible region and the undetermined region shrink. This is because the increase in the quantity of energy storage leads to a corresponding rise in the number of model parameters, making the selected physical model more refined and flexible. Therefore, we can conclude that as the number of the storage in the $\Omega_{phy}$ increases, the identification result is more accurate and consistent with the theoretical results.

The above results have some practical significance for real-world applications. First, after collecting the data, we should test if they contain sufficient information about the operational characteristics of the PFL by analyzing the set $\Delta\Omega$. Second, after obtaining the identification result of the PFL, we should check if the physical model $\Omega_{phy}$ is appropriately selected by inspecting if the practically infeasible region is included in the identification result.

*(3) Test with noisy data*

To provide a deeper understanding of the robustness and applicability of the model under noisy data, we perform the simulations using the datasets with noise. According to the requirements of China's national standard for the accuracy of electric meters, we assume that the error of the measured electric power is ±0.5% of the true value. We assume that the electricity price has a relative error of ±10%. We use Gaussian white noise in the simulations to model the errors based on these settings. Specifically, for the aggregate power $P_*^k$, we assume that the relative error $e_P$ obeys a Gaussian distribution with a mean of 0 and a standard deviation of $\sigma_P$, i.e., $e_P \sim \mathcal{N}(0, \sigma_P^2)$. We adopt the $3\sigma$ criterion, i.e., the probability that the noise value is distributed in $(-3\sigma_P, +3\sigma_P)$ is 99.74%, and hence $\sigma_P$ is set to (0.5/3)%. For the electricity price $\lambda^k$, we assume that the relative error $e_\lambda$ obeys a Gaussian distribution with a mean of 0 and a standard deviation of $\sigma_\lambda$, i.e., $e_\lambda \sim N(0, \sigma_\lambda^2)$, in which $\sigma_\lambda$ is set to (10/3)% based on the $3\sigma$ criterion. Based on the above settings, the original noise-free dataset $(\lambda^k, P_*^k), \forall k \in \mathrm{K}$ turns to the new noisy dataset, denoted as $(\tilde{\lambda}^k, \tilde{P}_*^k), \forall k \in \mathrm{K}$. Note that since the aggregated power is calculated based on the response model (1), we first obtain $\tilde{\lambda}^k = (1 + e_\lambda)\lambda^k$, and then calculate the aggregated power $P_*^k(\tilde{\lambda}^k)$ using the model (1) under the noisy electricity price $\tilde{\lambda}^k, \forall k \in \mathrm{K}$, and then obtain the noisy aggregated power $\tilde{P}_*^k$ by adding the noise to $P_*^k(\tilde{\lambda}^k)$, i.e., $\tilde{P}_*^k = (1 + e_P)P_*^k(\tilde{\lambda}^k)$. We use 50 samples, the same as those in the noise-free case in the following test.

First, we investigate the impact of the noise on the set $\mathrm{Conv}(\Gamma)$, as given in Fig. E-2 (a). The results under noise-free and noisy data are subscripted with $de$ and $un$, respectively. It can be observed that there are some minor differences between the sets $\mathrm{Conv}(\Gamma_{de})$ and $\mathrm{Conv}(\Gamma_{un})$. Interestingly, the set $\mathrm{Conv}(\Gamma_{un})$ completely covers the set $\mathrm{Conv}(\Gamma_{de})$. The reason is given as follows: (1) The response of the PFL, as shown in the model (1), depends on the direction of the electricity price $\tilde{\lambda}^k$, and the noise $e_\lambda$ in this case does not significantly change the direction of the electricity price $\lambda^k$; (2) Hence, the aggregated power $P_*^k(\tilde{\lambda}^k)$ calculated under the noisy electricity price $\tilde{\lambda}^k$ almost overlapping with $P_*^k(\lambda^k)$; (3) After adding the noise to $P_*^k(\tilde{\lambda}^k)$, the points on the boundary of $\mathrm{Conv}(\Gamma_{de})$ have 50% probability of falling outside the boundary $\partial\mathrm{Conv}(\Gamma_{de})$ since the errors are symmetric, making the set $\mathrm{Conv}(\Gamma_{un})$ with $\tilde{P}_*^k$ as the extreme points cover the set $\mathrm{Conv}(\Gamma_{de})$. This result reveals that in practical use, it is crucial to accurately predict the relative magnitude of electricity prices in different time periods. Besides, under noisy data, the region B is included in the set $\mathrm{Conv}(\Gamma_{un})$, which is easily mistaken as part of the feasible region of the PFL.

Second, we investigate the impact of the noise on sets $\Pi$. It is easy to notice that the shape of $\Pi$ has changed considerably by the noise. The reason is that the direction of the boundary $\partial\Pi$, as shown in Fig. E-2 (a), is determined by the direction of the electricity price, which is changed by the noise. Note that the change in the shape of $\Pi$ is not essential, and what we are concerned about is the operational characteristics of the PFL. In the results under noise-free data, the region A is an undetermined region since it is out of the set $\mathrm{Conv}(\Gamma_{de})$ but bounded by the boundary $\partial\Pi_{de}$. Nevertheless, it is identified as the infeasible region of the PFL under noisy data since it is not within the boundary $\partial\Pi_{un}$. Moreover, the regions B and C are identified by using noise-free data as infeasible regions since they are not within the boundary $\partial\Pi_{de}$. Besides, it also easy to mistake the region C as an undetermined region only based on the evidence that it is bounded by $\partial\Pi_{un}$. In summary, we can conclude that, under noisy data, if we use the results in Fig. E-2 (a) to prejudge the operational characteristics of the PFL as what we do under the noise-free dataset, we will get some wrong results, including misidentifying the regions that are actually feasible as infeasible regions, and mistakenly identifying the actually infeasible areas as feasible or undetermined regions.

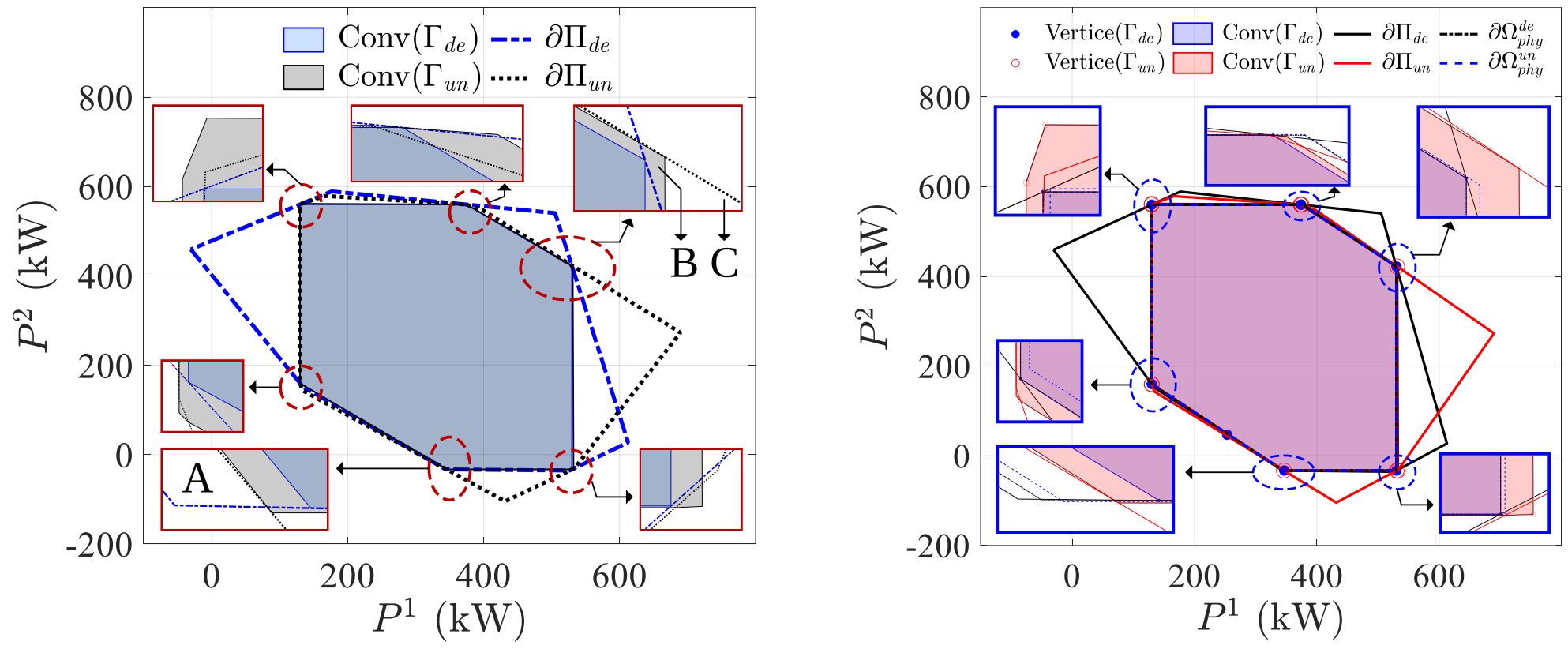

(a) (b)

Fig. E-2. (a) $\mathrm{Conv}(\Gamma)$ and $\Pi$ under noise-free data and noisy data, respectively; (b) Identification results of $\Omega_{phy}$ under noise-free data and noisy data (noise-free data for $\Omega_{phy}^{de}$ and noisy data for $\Omega_{phy}^{un}$). (A: Undetermined region under noise-free data but infeasible region under noisy data; B: Infeasible region under noise-free data but feasible region under noisy data; C: Infeasible region under noise-free data but undetermined region under noisy data.)

Third, we analyze the identification results. The feasible regions of the aggregated power in $\Omega_{phy}^{de}$ and $\Omega_{phy}^{un}$ are given in Fig. E-2 (b). In addition to the similar conclusions that we have previously obtained, we also find that there is very little difference between $\Omega_{phy}^{de}$ and $\Omega_{phy}^{un}$ either in shape or in the size of the regions. This means that a certain range of noise on the dataset has little effect on the identification results, although both the set $\mathrm{Conv}(\Gamma)$ and the set $\Pi$ are considerably affected by noise. That is, the physical model identified using the actual noisy data is very close to the one identified using the noise-free data. A potential reason is that the prior physical model, $\Omega_{phy}^{de}$ or $\Omega_{phy}^{un}$, endow the PFL model with a specific structural characteristic, making it less sensitive to the noise in the dataset. This result indicates that in practical use, embedding the prior physical knowledge into the data-driven modeling of the PFL greatly improves the robustness against the noise.

*(4) Conclusions*

The above results validate the effectiveness of theoretical results. Some conclusions drawn from the simulation results can be summarized as follows.

1) When $\Delta\Omega \neq \emptyset$, there always exist some undetermined regions, which we cannot use data alone to ensure whether they are part of the feasible region of the PFL.

2) The identification results under noisy data could misidentify the regions that are actually feasible as infeasible regions, or mistakenly identify the actually infeasible areas as feasible or undetermined regions.

3) If a priori physical model $\Omega_{phy}$ is appropriately selected, allowing a certain range of noise on the dataset has little effect on the identification results, although both the set $\mathrm{Conv}(\Gamma)$ and the set $\Pi$ are affected by noise.

Based on the simulation results, we can also get some practical implications for the data-driven modeling of the PFL in the real world, as follows.

1) After collecting the data, we should test if they contain sufficient information about the operational characteristics of the PFL by analyzing the set $\Delta\Omega$. The information of the undetermined region can be used to design the probing price to eliminate the information gap between $\mathrm{Conv}(\Gamma)$ and $\Pi$, and thus ensure the practical identifiability of the PFL.

2) Selecting an appropriate physical model $\Omega_{phy}$ will help identify the PFL model under incomplete and noisy information, while an incorrect $\Omega_{phy}$ produces wrong results. After obtaining the identification result of the PFL, we should check if the physical model $\Omega_{phy}$ is appropriately selected by inspecting if any response power newly measured is out of the feasible range calculated by the identification model.

3) It is crucial to accurately predict the relative magnitude of electricity prices in different time periods because the response of the PFL depends on the direction of the electricity price vector.